# Kondo effect and spin filtering in coupled quantum dots

S. Lipiński* and D. Krychowski

Institute of Molecular Physics, Polish Academy of Sciences, ul. Smoluchowskiego17, 60-179 Poznań, Poland

**Abstract**

The coherent transport through a set of N quantum dots coupled in parallel is considered in the limit of infinite intradot and finite or infinite interdot interactions. The mean field slave boson approach and the equation of motion method are used. For the full spin-orbit degenerate case the low energy behavior is characterized by an SU(2N) symmetry with entangled spin and charge correlations. The magnetic field breaks the spin degeneracy, but for the special choices of gate voltages the degeneracy might be recovered in one spin channel allowing the spin filtering.

## 1. Introduction

Spin dependent electron tunneling through nanostructures is of current interest because of its potential applications in field sensors, magnetic storage technology [1] and in quantum computers [2]. Quantum dot devices provide a well-controlled object for studying quantum many-body physics. One of the paradigmatic phenomena in the physics of strongly correlated electrons is the Kondo effect observed in metallic [3], semiconductor [4] and molecular nanostructures [5]. Apart from the spin degeneracy also orbital [6,7] or charge degeneracy [8-11] can play the role in this phenomena. The interplay of different degrees of freedom leads to a formation of highly entangled many body state.

Our present study is motivated by the observation of Kondo effect in electrostatically coupled quantum dots [8,9], triangular artificial atoms [6], and in carbon nanotubes [7]. The large number of tunable parameters in these systems allows the delicate manipulation of the Kondo physics. In this paper, we discuss the influence of interdot interaction including also the attraction case and examine how the picture is modified by the interdot tunneling. For infinite interdot interaction and fully symmetric case the conductance of the coupled dot system decreases with the increase of the number of coupled dots. The increased degeneracy yields also an enhancement of Kondo temperature. Partial breaking of the degeneracy by magnetic field or by the difference of gate voltages applied to the dots results in a crossover to Kondo physics in spin or charge sectors. For the case of orbital degeneracy in the one spin channel the system can operate as a spin filter.

*Corresponding author e-mail: lipinski@ifmpan.poznan.pl

## 2. Model

We consider the system of $N$ quantum dots coupled in parallel. Each of the dots is connected to a separate pair of electrodes. The system is modeled by the N-site Anderson Hamiltonian with additional interdot interaction and tunnel coupling terms:

$$H = \sum_{k\alpha i\sigma} \varepsilon_{k\alpha i\sigma} c^+_{k\alpha i\sigma} c_{k\alpha i\sigma} + \sum_{i\sigma} \varepsilon_{i\sigma} d^+_{i\sigma} d_{i\sigma} + t_0 \sum_{k\alpha i\sigma} \left( c^+_{k\alpha i\sigma} d_{i\sigma} + h.c \right) + t \sum_{i<j,\sigma} \left( d^+_{i\sigma} d_{j\sigma} + h.c. \right)$$
$$+ U_0 \sum_i n_{i+} n_{i-} + U \sum_{i<j,\sigma\sigma'} n_{i\sigma} n_{j\sigma'} \qquad (1)$$

where i numbers the dots, the leads are labeled by ($i,\alpha$), $\alpha=L, R$. $\varepsilon_{i\sigma} = \varepsilon_i + g_i\sigma h$, $\sigma = \pm 1$ (we set $|e| = \mu_B = 1$). The first term describes electrons in the electrodes, the second represents the field dependent dot energies, the third and fourth terms describe the tunneling between the electrodes and the dots and between the dots and the last two terms account for the site Coulomb and intersite interactions. The intradot charging energy is assumed to be strong ($U_0 \to \infty$) and we discuss the two opposite limits of weak ($|U| \sim \Gamma$) and strong ($U \to \infty$) interdot interaction. $\Gamma$ denotes the coupling strength to the electrodes $\Gamma(\omega) = \pi t_0^2 \Sigma_{k\alpha\sigma} \delta(\varepsilon - \varepsilon_{k\alpha\sigma})$, which for the assumed rectangular density of states is constant ($\Gamma = \Gamma(\omega)$) and is taken as the energy unit.

## 2. Results

### A. Weak dot-dot interaction

In this case each dot is simple occupied $<n_i> \approx 1$ and behaves as a separate magnetic (Kondo) impurity. The Hamiltonian may be written in terms of slave boson operators $b_i^+$, which create the empty states and pseudofermion operators $f_{i\sigma}$, which annihilate the singly occupied states [12]. The model is analyzed using the slave-boson mean field approximation (SBMFA), in which boson fields are replaced by their mean values. In the case of symmetric dots ($\varepsilon_{i\sigma} = \varepsilon_{j\sigma}$, $<b_i^+> \equiv b$) the SBMFA Hamiltonian reads:

$$H = \sum_{k\alpha i\sigma} \varepsilon_{k\alpha i\sigma} c^+_{k\alpha i\sigma} c_{k\alpha i\sigma} + \sum_{i\sigma} \varepsilon_{i\sigma} f^+_{i\sigma} f_{i\sigma} + t_0 b \sum_{k\alpha i\sigma} \left( c^+_{k\alpha i\sigma} f_{i\sigma} + h.c. \right) + t b^2 \sum_{i<j,\sigma} \left( f^+_{i\sigma} f_{j\sigma} + h.c. \right)$$
$$+ \sum_i \lambda \left( \sum_\sigma f^+_{i\sigma} f_{i\sigma} + b^2 - 1 \right) + \binom{N}{2} U \left(1 - b^2\right)^2 \quad (2)$$

The term with Lagrange multiplier $\lambda$ prevents double occupancy of the dot. The parameters $b$ and $\lambda$ are determined minimizing the ground state energy of (2).

The conductance expressed by a Landauer-type formula generalized to interacting system reads [13]:

$$g = \sum_{i\sigma} g_{i\sigma} = \frac{e^2}{h} b^2 \sum_{i\sigma} \int_{-\infty}^{\infty} d\omega f'(\omega) \text{Im}\left[G_{i\sigma,i\sigma}(\omega)\right] \quad (3)$$

where the dot Green's function is:

$$G_{i\sigma,i\sigma} = \cfrac{1}{\omega - \tilde{\varepsilon}_{i\sigma} + i\tilde{\Gamma} - \cfrac{\tilde{t}^2}{\omega - \tilde{\varepsilon}_{i\sigma} + i\tilde{\Gamma}}} \quad (4)$$

and $\tilde{\Gamma} = b^2 \Gamma$, $\tilde{t} = tb^2$. The renormalized energy $\tilde{\varepsilon}_{i\sigma} = \varepsilon_{i\sigma} + \lambda$.

The assumed value of the half of the bandwidth in the calculations was $D = 50$.

Fig. 1 presents conductance of the double dot system (DQD) ($N = 2$) for vanishing interdot tunneling ($t = 0$) and different values of interdot interaction $U$.

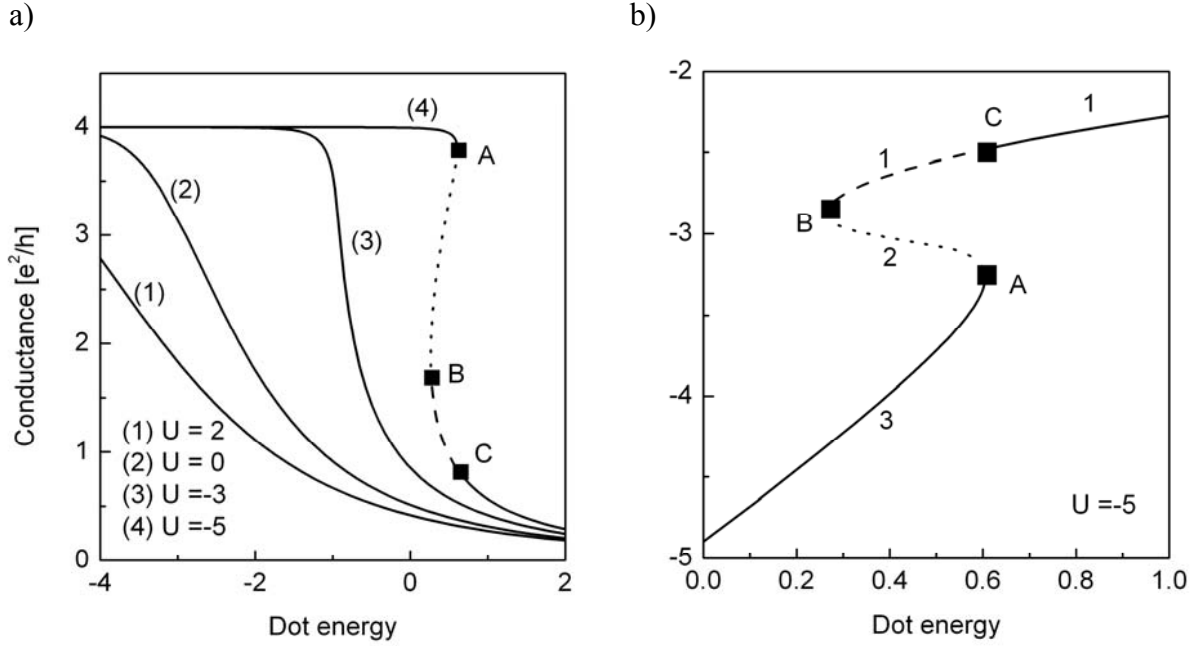

Fig.1 (a) Conductance of the symmetric DQD vs. dot energy $\varepsilon = \varepsilon_1 = \varepsilon_2$ for different values of interdot interactions U. (b) Energies of the solutions of SBMFA equations for attractive interdot interaction $U = -5$. Squares on the lines are only guides for eye marking the range of solutions.

The intersite repulsion ($U > 0$) raises the dot levels $\varepsilon_{i\sigma}$ effectively. As a result, the valence fluctuating regime is extended and the Kondo region shifts to lower values of $\varepsilon_{i\sigma}$. For negative values of U the opposite tendency is observed. Discussion of negative $U$ range is addressed to these molecular systems, in which the coupling with the lattice or with other quasibosonic excitations can lead to intersite attraction strong enough to dominate over the intersite Coulomb repulsion [14]. For stronger attraction ($-U/\Gamma > 3.75$) the self-consistent slave boson MFA equations might have three solutions (Fig 1b). At some value of site energy the mixed valence solution (line1) becomes metastable and the system abruptly relaxes to the new (Kondo) ground state (line 3) producing a discontinuous behavior of the current. The rapid formation of Kondo correlation with lowering the site energies is succored by attractive interaction favoring maximal occupancy of the dot ($n = 1$). The observation of multiple solutions should be taken with caution and requires checking whether it is not an artifact of the method (MFA). The mean field approximation neglects charge fluctuations, which are of importance in this region. The preliminary results obtained by the equation of motion method confirm the possibility of multiple solutions, but certainly more advanced methods are very much required to give an unambiguous answer.

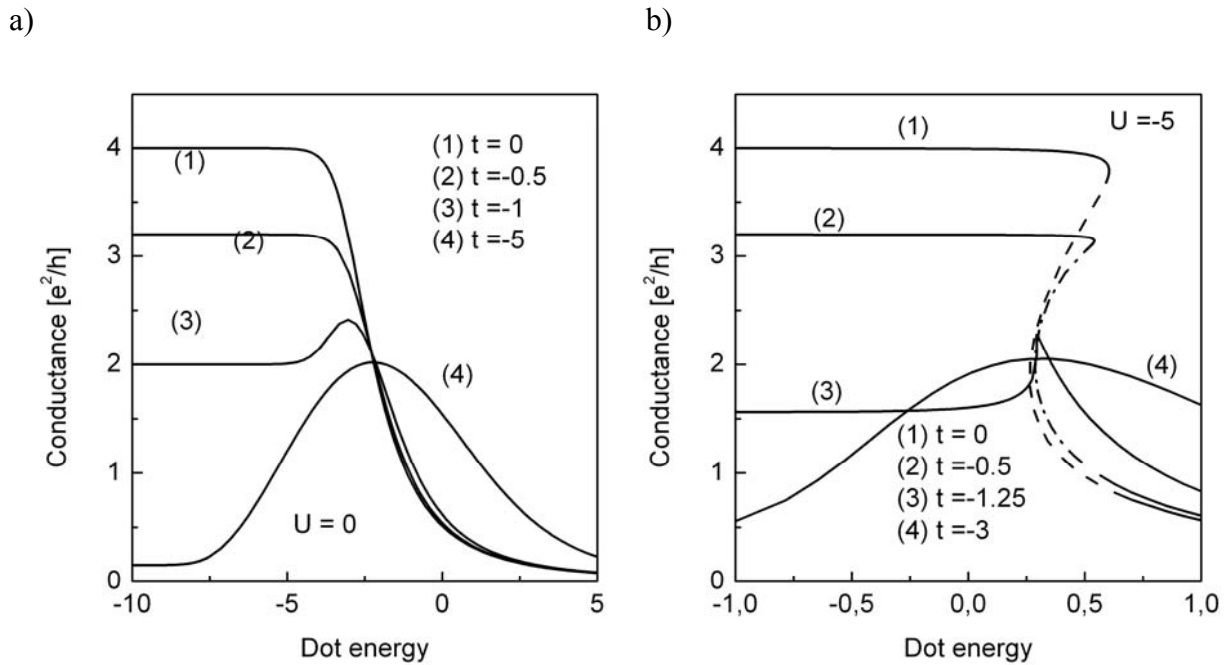

Fig.2 Conductance of the symmetric DQD vs. dot energy for different values of interdot tunneling: (a) $U = 0$, (b) $U = -5$.

Fig. 2 presents the influence of direct tunneling on the conductance for $U=0$ and $U=-5$. The result is similar to what is known in literature for the dots in series [15]. The resonant state has two peaks below and above the Fermi level for $|t|/\Gamma > 1$, whereas it has a single peak at the Fermi level for $|t|/\Gamma < 1$. In consequence a drop of conductance is observed for deeper site energies. For $U < 0$ the earlier mentioned enhancement of Kondo effect competes with the suppression caused by interdot tunneling, what is seen on Fig. 2b.

**B. Strong dot-dot interaction ($U \to \infty$)**

In this case only one dot can be charged at a given time. It is enough to introduce one boson field and one constraint that preserves the condition $\Sigma_i \langle n_i \rangle = 1$. The rest of the calculations follows the lines outlined in Section A. We restrict in the following to the case of vanishing interdot tunneling ($t=0$). For equal site energies of the dots $\varepsilon_i = \varepsilon$ all $N$ charge states $\{n_i = 1, n_j = 0, j \neq i\}$ are degenerate.

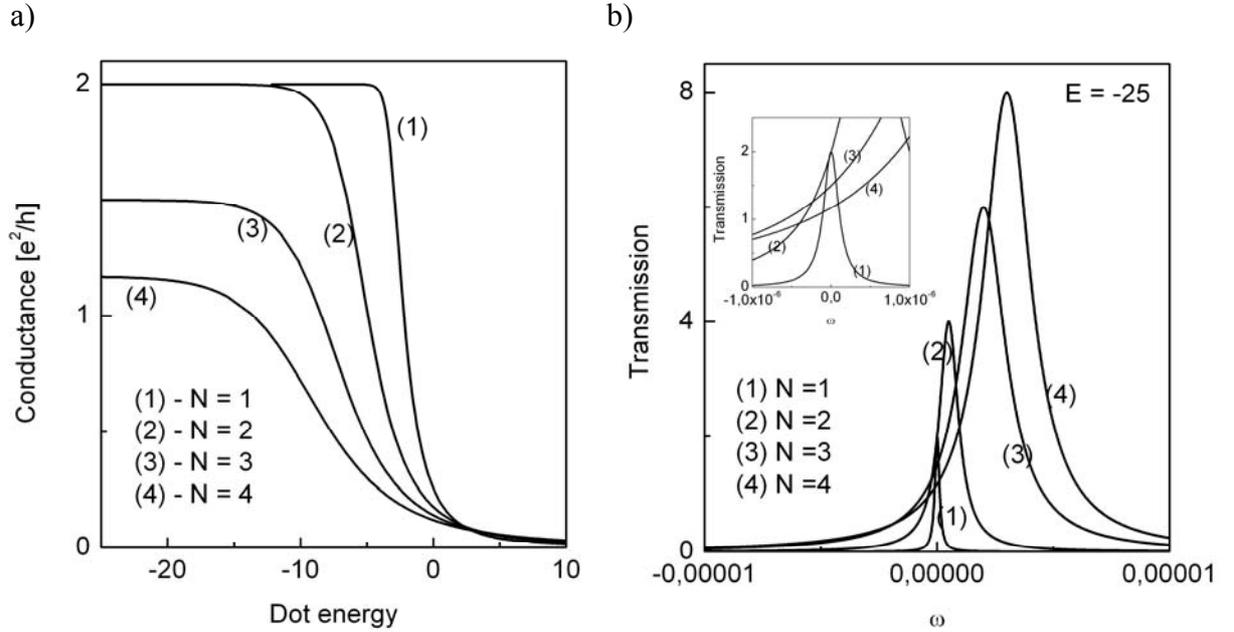

Fig.3 (a) Conductance of $N$ capacitively coupled quantum dots (NQD) for the fully degenerate case ($\Delta E = \varepsilon_2 - \varepsilon_1 = 0$, $h = 0$) in the limit of infinite intra and interdot interactions. (b) The examples of transmission of NQD for dot energy $\varepsilon = -25$. Inset shows transmission close to the Fermi energy.

For vanishing magnetic field $h = 0$ Kondo effect appears simultaneously in both spin and charge sectors resulting in an SU(2N) Fermi liquid ground state with a phase shift $\pi/2N$. For the deep dot levels conduction reaches the values $(2e^2/h) N \sin^2(\pi/2N)$ (Fig. 3a). The examples of the transmission function $T(\omega) = b^2 \Sigma_{i\sigma} Im[G_{i\sigma,i\sigma}(\omega)]$ are shown on Fig. 3b. From the position of the peaks or from their widths it can be inferred that the increased degeneracy yields an enhancement of $T_k$, which makes the observation of strong entanglement of charge and spin flips events more accessible. In the following we discuss the polarization of conductance defined by $PC(i) = (g_{i+} - g_{i-}) / (g_{i+} + g_{i-})$. Our further discussion is devoted to the double dot system with different site energies $\Delta E \equiv (\varepsilon_2 - \varepsilon_1) \neq 0$, placed in a magnetic field $h \neq 0$. In general the degeneracy is removed and Kondo effect disappears. By an appropriate tuning of gate voltages however the orbital degeneracy can be recovered (SU(2)). For $g_1 \neq g_2$ both types of orbital Kondo effects occur: the spin conserving effect ($\varepsilon_{1+} = \varepsilon_{2+}$ or $\varepsilon_{1-} = \varepsilon_{2-}$) and spin mixing ($\varepsilon_{1\sigma} = \varepsilon_{2-\sigma}$) [16]. Both types of effects can be used for spin filtering. Here we discuss the cases $|g_1| = |g_2|$, where only one type of Kondo effect occurs. For $g_1 = -g_2$ only one spin channel contributes to the Kondo resonance, for $g_1 = g_2$ the orbital isospin fluctuations mix different spin channels. The corresponding spin filtering is shown on Fig. 4a. The results were obtained by EOM using the decoupling procedure proposed by Lacroix [17]. For more details see [16]. For $g_1 = -g_2$ the preferred spin direction for tunneling through the dots is the same for both of them, for $g_1 = g_2$ the signs of polarization of conductance are opposite. The values of $PC$ presented on Fig. 4a correspond to the maximum of $PC(h)$ curves and they occur for $h = \pm \Delta E/2$ (Fig. 4b). For small values of $\Delta E$ i.e. closer to the full degeneracy point larger number of processes contribute to the many body structure of the densities of states at the Fermi level. In some cases it results in the change of proportion of spin up and spin down contributions to DOS in this region, what in turn leads to the reversing of the sign of polarization of conductance. It is seen for example for dot 1 (QD1) in the low $\Delta E$ region (Figures 4a and 4c).

To summarize, we have shown that the intersite repulsion suppresses Kondo effect and intersite attraction enhances it. The crude MFA results suggest a discontinuous transition between the mixed valence and Kondo range for negative U. But the tendency to a jump is

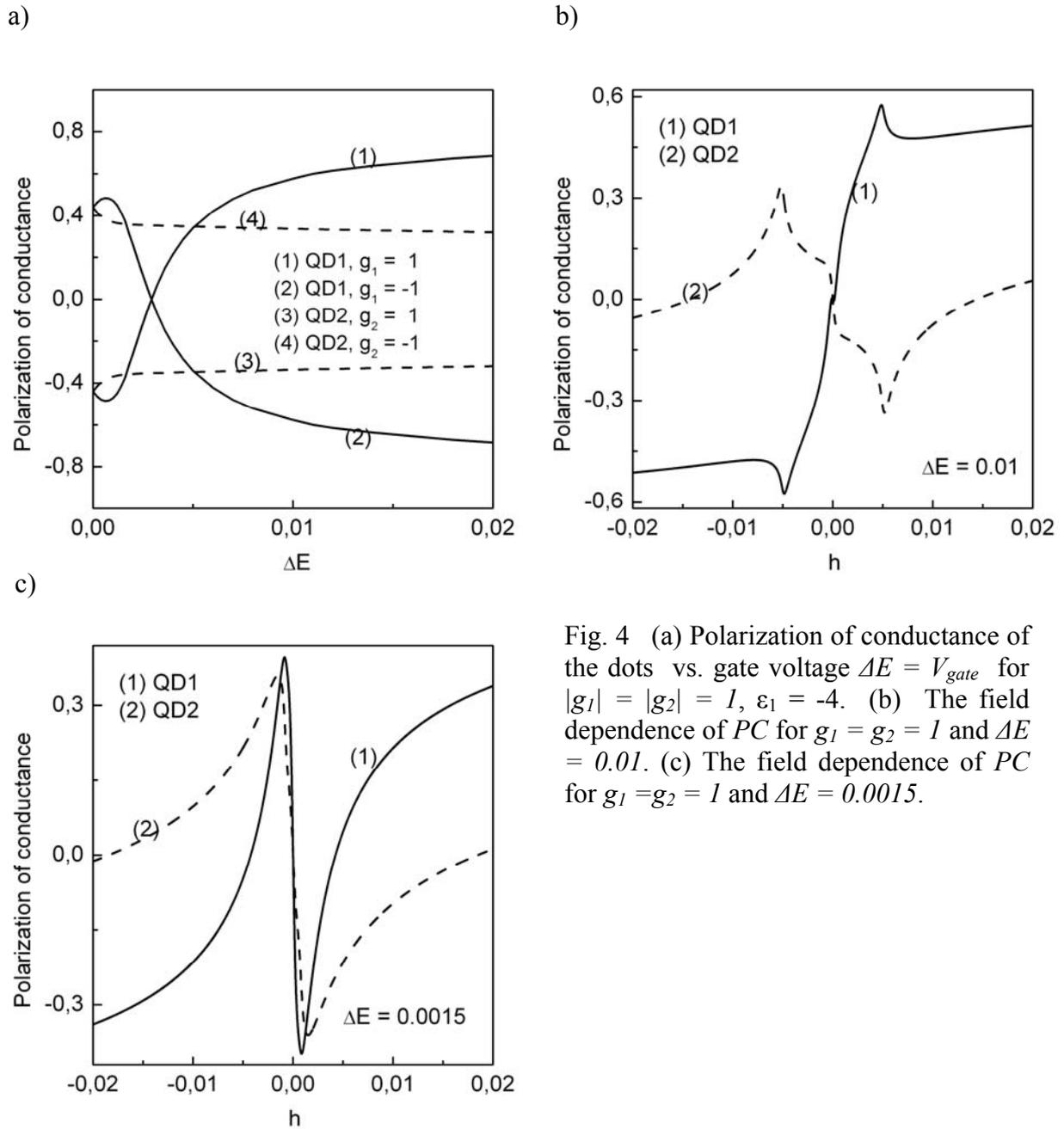

Fig. 4 (a) Polarization of conductance of the dots vs. gate voltage $\Delta E = V_{gate}$ for $|g_1| = |g_2| = 1$, $\varepsilon_1 = -4$. (b) The field dependence of PC for $g_1 = g_2 = 1$ and $\Delta E = 0.01$. (c) The field dependence of PC for $g_1 = g_2 = 1$ and $\Delta E = 0.0015$.

perhaps somewhat exaggerated due to neglect of fluctuations. For real systems, where also direct tunneling is of importance, one can expect smooth transition even in the simple MFA picture. Tunneling rearranges the effective energy level and suppresses Kondo effect. We have also presented, that the increase of orbital degeneracy caused by coupling of a larger number of the dots enhances Kondo temperature. Finally we have shown how the high spin-polarized transmission of the same sign in both dots or of opposite signs can result in the orbital Kondo range.

**Acknowledgements**   The work has been supported by the Polish State Committee for Scientific Research through the project PBZ/KBN/PO3/2001 and by MAG-El-MAT network.